**Quantum Brain Dynamics. A Possibility of Having a Quantum Interpretation of the Brain**

Dozie Iwuh O.S.A
Augustinian Institute in Makurdi, an affiliate of the Benue State University Makurdi, Benue State.
Email: registered501@gmail.com
Phone: 08030635406

**Abstract**

*In recent years we have seen quantum physics advance in leaps and bounds. Living matter is becoming more comprehensible when we appeal to the fundamental states that define their existence, namely the quantum fields that comprise them. When we consider living matter (organisms), we are presented with the difficult complexity of understanding the human brain. The brain itself is not the human being neither does it comprise the totality of the human person, but is inherently embedded in the entire being of the human person. We are obviously aware of the fact that the major function of the brain pertains to consciousness, be it broad and specific, the former being that general/objective view of awareness recognized, such that is seen when one wakes up from sleep, and the specific that refers to the particular/subjective state of being aware of this or that (this comes after the broad though). Other functions include memory, thought-control, motor-skills, vision, breathing, temperature, body-regulation etc. However, all these come after consciousness. Thus what Quantum Brain Dynamics (QBD) considers is not just these other functions of the brain, this is because they can be well analyzed with the workings of classical mechanics (even though they still play host to a quantum description). It rather considers two specific functions above all else consciousness and memory. QBD falls in line umbrella-covers aspects of the quantum brain analysis such as quantum-consciousness, quantum-mind and quantum-brain. The inspiration that lurks behind the Quantum Interpretation of the Brain (QIB), is traceable to the 1944 article written by E. Schrodinger, What is Life, in which he presents how a living organisms evades decay to equilibrium by the fact of negentropy, as such life which is in its ordered macroscopic state is created (in an environment of disorder), which moves against the second law of thermodynamics. The life that is created, that which is sustained, arises from an interaction that the organism engages in with the environment. This interaction is microscopic, albeit quantum, it is an interaction that underscores the reality of quantum entanglement (which also plays hosts to the superposition of quantum states). The quantum interpretation of the brain is a nascent, yet burgeoning as it might be that necessary tool required for a better articulation and comprehension of the brain.*

*Key words: Brain, Quantum states, Superposition, Entanglement, Memory*

**What is Quantum Brain Dynamics**

We do not seek out a formal definition of Quantum Brain Dynamics, but a description. In the light of the aforementioned, thus, Quantum Brain Dynamics (QBD) describes long range ordered dynamics of the quantum system of electromagnetic field and water dipole field in the brain.[1] QBD is nothing else but Quantum Electrodynamics (QED) of the electric dipole field of dipolar solitons and water molecules with a symmetry property under the dipole rotation. The highly systematized functioning of the brain, according to QBD, is found to be realized by the spontaneous symmetry breaking (SBS) phenomena. Memory printing, recall and decay processes are represented by the fundamental physical processes standing for the phase transition process, the symmetry restoring process and the quantum tunneling process, respectively.[2] The brain as already noted is a typical macroscopic object extraordinary in its functioning as it gives rise to highly advanced mental "objects" such as consciousness (plus the unconscious), mind, memory and will (including all that pertains to the human cognitive abilities). It is the custom and cognitive sciences to regard the brain as a tissue made of a huge of brain cells, and many phenomenological theories of brain functioning based on the macroscopic picture of electric and chemical circuits of cells take into account various mesoscopic aspects of the brain cell revealed by molecular biological studies.[3] Of all living matter, the brain is considered as the highest example. Therefore, an investigation of the brain and its functionality pertains to the Quantum field, which QBD, as proposed in this paper, is well equipped with affording us a better comprehension.

---

[1] M. Jibu – K. Yasue, *Magic without Magic. Meaning of Quantum Brain Dynamics* in the Journal of Mind and Behaviour, Vol 18, No2/3, 205-227. This is proposed as a revival of the original idea that was developed by H. Umezawa in the early 1960s.

QBD is a completely new theoretical framework to the description of the fundamental physical process of the brain dynamics that makes up the human on the basis of quantum field theoretical analysis of the fundamental system of brain tissue.[4] What should be referenced always is that "the brain is essentially a microscopic quantum dynamical system with macroscopic spatial extent that is quantum field dynamics".[5] Which means that the functioning of the brain is more quantum that it is classical, more quantum than macroscopic. QBD, derives its mathematical underpinnings from quantum field theory or quantum field dynamics. Thus we can in extension say that the brain can be analyzed and understood with excerpts gathered from quantum field theory or dynamics. We should also note that because the brain "is a mixed physical system of quantum dynamical system and classical dynamical system",[6] QBD is an intersection between the quantum and the classical mode.

In its nature, QBD is dissipative, this owes a lot to the fact that it is carried out in complex systems that are open. According to Y. Bar-Yam,

> A complex system is a system formed out of many components whose behavior is emergent that is, the behavior of the system cannot be simply inferred from the behavior of its components. The amount of information necessary to describe the behavior of such a system is a measure of its complexity.[7]

We can yet compute a complex system to be "an ensemble of many elements which are interacting in a disordered way resulting in robust organization and memory".[8] Quantum systems are systems that interact with their environment; they are characterized by "these interactions and properties

---

of the environment".[9] This implies that quantum systems are open systems. Open systems are to be contrasted with isolated and closed systems, for the open systems exchange energy and matter, while the closed, exchange only energy, and in the isolated, there is no exchange (this is actually in an idealistic reality). With this said, when it comes to computing the total energy, the Hamiltonian, of the system, the open system, it becomes an arduous task; this is because the arena of experimentation is not controlled. This means that might not really be able the next flow or net exchange of the materials involved, which leaves us at best with a not to fair approximation of the total energy. Therefore to overcome this challenge, the need to close the system arises. It should be recalled that the system being considered here is the dissipative open system. In trying to compute this, we

> must include the details of the processes responsible for dissipation: thus the total Hamiltonian must describe the system, the bath and the system-bath interaction. It turns out that the canonical commutation relations (CCR) are not preserved by time evolution due to damping terms. By including the bath one "closes" the system in order to recover the canonical formalism and one realizes that the role of fluctuating forces is in fact the one of preserving the canonical structure of the CCR.[10]

Yet, this might prove to still be arduous, as "the knowledge of the details of the processes inducing the dissipation may not always be achievable",[11] thus what is done is that the degrees of freedom is doubled to attain to the requisite closure; eventually to do this, will imply that a mirror image of the system under consideration is created, which "behaves as a reservoir".[12]

QBD has been advanced by the findings of K. Pribram, K. Yasue, M. Jibu, G.G. Globus etc., but such advancement came on the heels of a proposal arrived at by H. Umezawa, who saw it needful to apply the findings of Quantum Field Theory QFT, to the brain and all that pertains to it, especially consciousness. It can be stated that the emergence of QBD arises from the application of QFT to biological processes. This is because the brain is a biological organism that plays host to evolution in its functions and operations. Biological processes are mainly,

> at (their) base(s), biochemical, and at a deeper level the laws of quantum physics still come into play. Currently many "solid" physicists look at quantum physics with the eyes of "solid state physics" and this is still the dominant thought. But the rule is simple: any equation which includes the Planck constant is a quantum physics equation. Looked at this way, photochemical reactions, electron transfer and ion interaction, getting a suntan, photosynthesis, the sense of sight and the breakdown of DNA under ultraviolet light are all unquestionably quantum physical reactions. But even if these equations and the Planck constant appear in school books, it is not emphasized that these are "photochemical quantum equations". Most recently, with the advancement of research on photosynthesis, on magnetic direction finding in animals (magnetoception[13]) and on the sense of smell, quantum biology has become an accepted subject.[14]

---

[9] V. Semin – F. Petruccione, *Dynamical and Thermodynamical Approaches to Open Quantum Systems* in Scientific Reports, 2020, 10, 2607, 1-10.
[10] M. Blasone, P. Jizba & G. Vitiello, *Quantum Field Theory and its Macroscopic Manifestations. Boson Condensation, Ordered Patterns and Topological Defects*, Imperial College Press, Convent-Garden, London, 2011, 373.
[11] Ibid.
[12] Ibid.

We can denote QBD to be the adaptation of the findings of QFT to the brain, as an expected hypothesis, that can and will be used to "describe the mechanism of memory in the brain".[15] For memory to be described, it means that consciousness must be contained, necessarily in that description. Again we should add that QBD is quantum electrodynamics (QED) of the dipole field with symmetry under the dipole rotation.[16]

**Mathematico-Physical Representation of QBD**[17]

It should be stated that one fundamental requirement in QFT is in the infinite number of degrees of freedom. In such wise in QBD, there are two molecules that are involved, namely, the water molecules and biomolecules. The degrees of freedom thus will include the molecular vibrational fields of the biomolecules and the water molecules (these vibrational fields are in an angular motion of spins). It was Frohlich that proposed that a theory that classed these "biomolecules with higher electric dipole moment that line up along the actin filaments immediately beneath the cell membrane, which propagate along each filament as coherent waves".[18] These coherent waves advance without the loss of heat. The neuron has always been and remains yet the most fundamental constituent of the brain that wades into all that pertains to consciousness and memory. Yet when it pertains to how memory is recalled or processed and consciousness arises, we realize that there is a movement from the internals of the neurons to the externals. That is to say that there is an interaction between neurons and even glial cells. Working with this reality, we can say that "the most fundamental unit of brain functioning, with its infinite degree of freedom, manifests spatial extension beyond a single cell".[19] According to Mari and Jibu,

> It is found that the principal degree of freedom of QBD, that is, the corticon is not merely the dipolar soliton as was expected at first sight but also the water dipole moment surrounding the protein filaments. The corticon in QBD is now fully described by the electric dipole field (of both dipolar solitons and water dipole moments) spanning the spatial volume of the brain tissue. In this sense, we may call the fundamental system of the brain… simply as "systems of corticons"…[20]

Since there is an infinite degree of freedom, it is supposed that there should necessarily exist a field entity, which can be described by QFT, and this field entity is called cortical field.[21] These cortical fields have an energy quanta referred to as corticons. These can exist both "Inside and outside neurons and glia everywhere in the cerebral cortex because the cortical fields extends through the whole cortex".[22] The corticons engage in interplay with the electric dipole field (EDF) in the brain. And this EDF comprises of the dipolar solitons and the water dipole moments. It is assumed thus that there is the existence of symmetry in the EDF under rotations that is to say that "even if the EDF on each position is rotated by any spatial angle, the total energy of the system of corticons is kept invariant".[23] Consequently, we can say that it is the creation and destruction of corticons (with its apparent interaction with the EDF) that gives rise to memory and consciousness.

---

[13] Magnetoception or magnetoreception refers to the ability of living things to sense direction, height and their position by the use of a magnetic field. It was first shown in 1966 in a study of European Robins that on the evidence of behaviour certain animals use the Earth's magnetic field to move over long or short distances, (W. Wiltschko – F.W. Merkel, *Orientierung zugunruhiger Rotkehlchen im statischen Magnetfeld* in Verh. dt. zool. Ges. 59, 1966, 362–367).
[14] S. Tarlaci – M. Pregnolato, *Quantum neurophysics: From non-living matter to quantum neurobiology and psychopathology* in International Journal of Psychophysiology 103, 2016, 161–173.
[15] A. Nishiyama et al., *Non-Equilibrium Quantum Brain Dynamics. Super-Radiance and Equilibration in 2 + 1 Dimensions* in Entropy, 21, 1066, 2019, 1-26.
[16] M. Jibu – K. Yasue, *Quantum Brain Dynamics and Quantum Field Theory*.
[17] This heading is important to draw our attention to the fact that what is being represented in QBD, is not the conventional mathematics, it is more of physics than it is mathematics.
[18] M. Jibu – K. Yasue, *Quantum Brain Dynamics and Consciousness. An Introduction*, John Benjamins Publishing Company, Amsterdam, 1995, 147.
[19] Ibid, 148.
[20] M. Jibu-K. Yasue, *Quantum Brain Dynamics and Quantum Field Theory*.
[21] M. Jibu-K. Yasue, *Quantum Brain Dynamics and Consciousness. An Introduction*, 148.

These vibrational fields of the biomolecules is referenced with the letter B and the latter referenced with the letter W.[24] Due to the fact that the vibrational fields of B, which will be now referred to as B-fields, come to existence from unlocalized electrons or holes that are contained in the protein filaments, it is right that we work with the two-component spinor field (that is to say that we are working with the spinor representation of the B-field).[25] The spinor representation is called to play in QFT because it is the "simplest way used to describe the electric dipole field is given by a spinor field".[26] It should be noted that a spinor is a two-component complex field that is usually described with a two by one matrix form, and in its representation, the electric dipole moments are elaborated by the Pauli spin matrices.[27] It ought to be noted that the spinor field does not stand for the electric dipole moment, but represents physically "the molecular vibrations of the protein filaments and the water molecules in the brain…"[28] We can as such denote the B-field to be:

$$\Psi_B(x, t) = \{\begin{matrix} \psi_B^+(x, t) \\ \psi_B^-(x, t) \end{matrix}\} \qquad (1)$$

The spinor components are represented with $\psi_B^+(x, t)$ and $\psi_B^-(x, t)$. The electric dipole moment of the B-field can be given as

$$\tilde{\psi}_B(x,t)\sigma\psi_B(x,t), \qquad (2)$$

we can produce an adjoint spinor from the above thus as

$$\tilde{\psi}_B(x,t)= (\psi_B^+(x, t)^* \;\; \psi_B^-(x, t)^*) \qquad (3)$$

The B-field is localized thus $\psi_B(x, t) \neq 0$ in a net mean position of $x = x_k$ of N protein filaments,[29] where k represents an integer value from 1 to N.

Hence from (2) above, we can write the localized electric polarization of the B-field as

$$T^k = \tilde{\psi}_B(x,t)\sigma\psi_B(x,t) \qquad (4)$$

As regards the W-field, we ought to recall that it comprises of the dipolar solitons and the water dipole moment, "it is a dipolar phonon field of water molecular vibrations".[30] We start off with the vector representation

---

[22] Ibid, 153.
[23] M. Jibu-K. Yasue, *Quantum Brain Dynamics and Quantum Field Theory*.
[24] M. Jibu - K. Yasue, *The Basics of Quantum Brain Dynamics*.
[25] Ibid.
[26] M. Jibu-K. Yasue, *Quantum Brain Dynamics and Quantum Field Theory*.
[27] Ibid.
[28] Ibid.
[29] M. Jibu – K. Yasue, *The Basics of Quantum Brain Dynamics*.
[30] Ibid.

$$R_w(x,t) = [R^1_w(x,t), R^2_w(x,t), R^3_w(x,t)\ldots R^n_w(x,t)] \quad (5)$$

Based on the quantization procedure in QFT, the Fourier transformation of the W-field is introduced such as

$$R_w(x,t) = 1/\sqrt{v}\{\sum_k[A_w(k,t)e^{ikx} + A_w^*(k,t)e^{-ikx}]\} \quad (6)$$

For the sake of specificity, we have to assume that the W-field manifests a polarization in the second direction. Thus we have

$$R_w(x,t) = 1/\sqrt{v}\{e_2\sum_k[A_w(k,t)e^{ikx} + A_w^*(k,t)e^{-ikx}]\} \quad (7)$$

Where $e_2$ represents the unit vector in the second direction.

We then assume that $A_w(k,t)$ and $A_w^*(k',t)$ are creation and annihilation operators that obey the canonical commutation relation $[q_x(t), p_y(t)] = i\delta_{xy}$ of

$$[A_w(k,t), A_w^*(k',t)] = \delta_{kk'} \quad (8)$$

Such that we have equal time commutation relations such as $[\varphi(x,t), \varphi(y,t)] = [\pi(x,t), \pi(y,t)] = 0$, but for QBD as

$$[A_w(k,t), A_w(k',t)] = [A_w^*(k,t), A_w^*(k',t)] = 0 \quad (9)$$

In QFT, we introduce canonical variables and the Hamiltonian operator; in this case we would be introducing the momentum operator (Q) and coordinate operators (P) that is related to the creation and annihilation operators thus:

$$A_W(k,t) = 1/\sqrt{2}\{\sqrt{K_k}Q_k(t) + i/\sqrt{K_k}(P_{-k}(t)\}$$
$$A^*_w(k,t) = 1/\sqrt{2}\{\sqrt{K_k}Q_{-k}(t) - i/\sqrt{K_k}(P_k(t)\} \quad (10)$$

Where $K_k$ is a constant that is inherent to the W-field

The canonical variables of the W-field should be able to satisfy the canonical commutation relation in QFT of $[P_k(t), Q_{k'}(t)] = -i\delta_{kk'}$ such that

$$[P_k(t), P_{k'}(t)] = [Q_k(t), Q_{k'}(t)] = 0$$
$$P^*_k = P_{-k}$$
$$Q^*_k = Q_{-k} \quad (11)$$

The free and interaction Hamiltonian of the B- and W-fields, which gives the total Hamiltonian can thus be given as

$$H_0 = \tfrac{1}{2}\{\sum_k[P^*_k(t), P_{k'}(t) + K^2_k Q^*_k(t), Q_{k'}(t)]\}$$
$$H' = {}^N\sum_{j=1}\sum_k(f/2\sqrt{V})\{T^j_1 Q_k(t)e^{ikxj} - T^j_2 P_k(t)/K_k(e^{ikxj})\} \quad (12)$$

Where f is the coupling constant between the B- and W-field

From equation (4) we see that the B-fields are subject to the commutation relation of

$$[T^j_1(t), T^j_2] = 2i\, T^j_3\, \delta_{jk}$$
$$[T^j_2(t), T^j_3] = 2i\, T^j_1\, \delta_{jk}$$
$$[T^j_3(t), T^j_1] = 2i\, T^j_2\, \delta_{jk} \quad (13)$$

As related to the canonical variables, $P_k(t)$; $Q_k(t)$ and $T^k(t) = \tilde{\psi}_B(x,t)\sigma\psi_B(x,t)$ and the total Hamiltonian that is derivable from an addition of the interacting Hamiltonian and the free

Hamiltonian H = H$_0$ + H', the quantum field dynamics of the B- and W-fields, takes on the Heisenberg's equation of

$$d/dt[(Q_k(t))] = -i[(Q_k(t), H]$$
$$d/dt[(P_k(t))] = -i[(P_k(t), H]$$
$$d/dt[(T^j(t))] = -i[(T^j(t), H] \quad (14)$$

It is thus noticed that the Hamiltonian remains invariant under the transformation of the canonical variables,[31] of

$$P_k(t) \rightarrow P^{\dagger}_k(t)$$
$$Q_k(t) \rightarrow Q^{\dagger}_k(t)$$
$$T^j_k(t) \rightarrow T^{J\dagger}_k(t) \quad (15)$$

**Does QBD Make The Quantum Interpretation Of The Brain Possible?**

The above mathematical rendition of QBD, already bespeaks of the fact that a quantum interpretation of the brain is already on-going. There are yet rough edges that needs to be straightened out, yet there is the optimism that this will occur in the very near future. In addition to this, the above mathematical representation, points to the fact that the quantum interpretation of the brain as articulated by QBD, is neither superstitious nor unfounded in the realm of quantum physics. The quantum interpretation of the brain, considers as already noted, memory and consciousness, amongst others functions of the brain that are analyzed. Nonetheless, this seems to be a necessity, because classical physics in respect to providing an explanation to the reality of consciousness, arrives at a problem known as the unity or binding problem. The binding problem is easily captured as: how can we account for the fact that our conscious experiences are united and not differentiated into parts? This is the reason that it is referred to as the unity problem.

As noted by V.G. Hardcastle,

> Our brains process visual data in segregated specialized cortical areas. As is commonly remarked the brain processes, the what and the where of its environment in separate, distal locations. Indeed, regarding the what information that the brain computes, it responds to edges, colours and movements using different neuronal pathways. Moreover, so far as we can tell, there are no true association areas in our cortices. There are no convergence zones where information is pooled and united; there are no central neural areas dedicated to information exchange. Still, the visual features that we extract separately have to come together in some way, since our experiences are of these features united together into a single unit. The binding problem is explaining how our brains do that, given the serial distributed nature of our visual processing.[32]

Another way to express the binding problem is that

> It concerns how visual system represents the hierarchical relationships between features (edges and objects). For example, at an object level, how does the visual system represent which low-level features belong to a particular object? If two letter T and L are seen

---

together, how does the visual system represent which horizontal and vertical bars are part of which letter?[33]

It can also be related thus, what is that property that controls, unifies, segregates all the physico-chemical processes taking place in the stratified society of brain cells?[34]

We first have to note that the unity or binding problem is one that pertains to consciousness, thus, with that being said, it thus is suggested that owing to the fact that the brain engages in electrochemical reactions fundamentally as regards the macroscopic functioning that is witnessed, QBD, that allows for a quantum interpretation, might just provide an answer. According to Jibu, Yasue and Hagan,

> The binding problem must be solved not by introducing the idealistic quantum mechanical nonlocality but by investigating the usually neglected quantum electromagnetic phenomena taking place in the dynamically ordered regions (…) of intracellular and extracellular water. There, each brain cell is enfolded within a common field of macroscopic condensation of evanescent photons and all the physico-chemical processes taking part in the stratified society of brain cells are subject to the control and unification by quantum electrodynamics. Unity of consciousness thus arises from the existence of the global field of condensed evanescent photons overlapping the whole brain tissue in the cranium.[35]

As regards the human memory, the shortcomings of classical physics and its attendant inquiry with the use of the microscope, falls short. This is coming on the trails of the fact that the brain is estimated to contain $10^{10}$ neurons with ten times as much glia cells which combines the neurons and with the cranial blood vessels. There is an apparent community of neuron networking in the brain as many, if not all, of the neurons, receive hundreds and thousands of fiber connections from other neurons.[36] This point just stated buttresses the earlier stated fact of the shortcoming of classical physics and its attendant use of the microscope in what pertains to an inquiry into the brain. Memory, long term memory in particular, comprehension and assimilation as found in learning and other cognitive functions that pertain to the brain, have been noted to arise from impulses that travel through these congregational net of neurons.[37] It has been experimentally proven that memory is not lost after ablation experiments or when it is sliced into many directions, even in the advent of the destruction of some neuronal networks.[38] This only goes to indicate the memory is not attached or wired strictly to neuronal parts of the brain, but "are rather diffused in the brain".[39] This occurs in two fronts, with the "first net that corresponds to an incoming message (that) has its information transferred to other nets far from the original location of the stimuli carrying the message".[40] Memory is diffused and not wired to any part of the brain, apart from being diffused, there is the phenomenon of printing and recalling as regards the memory and not just diffusion. This explanation is provided for by QBD.

---

[33] J.B. Isbister et al., *A New Approach to Solving the Feature- Binding Problem in Primate Vision* in Interface Focus 8, 2018, 1-23.
[34] M. Jibu-K. Yasue, *Quantum Brain Dynamics and Quantum Field Theory*.
[35] M. Jibu, K. Yasue and S. Hagan, *Evanscent (tunneling) photon and cellular 'vision'* in BioSystems, 42, 1997, 65-73.
[36] V. Braitenberg, et al., *Observations on Spike Sequences from Spontaneously Active Purkinje Cells in the Frog* in Kybernetik 2, 1965, 197-205.
[37] F.O. Schmitt, *Molecules and Memory* in New Scientist 23, 1964, 643-645.
[38] L.M. Ricciardi – H. Umezawa, *Brain and Physics of Many-Body Problems* in Brain and Being. *At the Boundary Between Science, Philosophy, Language and Arts*, G.G. Globus, K.H. Pribram and G. Vitiello (eds.), John Benjamins Publishing Company, Amsterdam, 2004, 257-268.
[39] F.O. Schmitt, *Molecules and Memory*.

Jibu and Yasue explains it thus:

> Let us start with the fundamental system of the brain, that is, the system of corticons where memory of a specific stimulus from the external world is yet to be printed. The brain tissue is here exposed to stray signals including unattached perception of external events as well as activity associated with general physiological events such as motor activity and the like. Those stray signals are organized and transmitted to the system of corticons through the metabolizing system (i.e., the neuronal network). Namely, they can create corticons indirectly in the fundamental system, and various spatial domains of the macroscopic ordered states are formed provided that the created corticons manifest a long-range correlation whose spatial extent is larger than the coherence length. Thus, there exist apparent thresholds for the incoming energy from the metabolizing system to the fundamental one to create ordered domains. The stray signals to the brain tissue transmitted with energy slightly exceeding the threshold value are all tacitly coded in the dynamical domain structure of macroscopic ordered states in the small domains. Notice that in such a dynamical structure the direction of electric dipole moments of the corticon field can randomly vary from domain to domain and there exists no hierarchy among the coded signal… Once memory of a typical external stimulus is printed in the macroscopic ordered state of the system of corticons, it can be recalled quite easily thanks to the Goldstone bosons.[41]

As regarding learning, of what memory and recall plays an integral role, Jibu and Yasue also notes:

> The learning process can be identified with the phase transition of the system of corticons from the less ordered state of many but small ordered domains to the more ordered state of a few but large ordered domains. Such a phase transition is induced by an external stimulus which supplies the system of corticons with enough energy to break domain boundaries of many small ordered domains, thus aligning the electric dipole moments of the corticon field in much larger domains. Of course, the notion of external stimulus denotes an energy flow organized and transmitted through the metabolizing system… By… a learning process, a typical external stimulus can be printed in the system of corticons of the fundamental system as a stable macroscopic ordered state. The stability of printed memory comes from the very fact that it is coded into the vacuum state of the corticon field the stability of which is a consequence of quantum field theory. Apparently, printed memory manifests nonlocal (i.e., diffused) existence.[42]

---

[40] L.M. Ricciardi – H. Umezawa, *Brain and Physics of Many-Body Problems*.
[41] M. Jibu-K. Yasue, *Quantum Brain Dynamics and Quantum Field Theory*. Nambu-Goldstone (NG) Bosons are massless scalar quanta (bosons) that arise when symmetry is spontaneously broken. These NG quanta propagate through the system and are the carriers of the ordering information. Due to the masslessness of the NGB, their condensation does not necessitate a change of energy state of the system, this is because when in the lowest state, the NG quanta does not carry energy. This is needed to enable the NG quanta cut across the full system volume, sending long distance informational correlation in the system, thus setting up an ordered pattern, that will enable transfer or diffusion or non-localization of information via long distances. (G. Vitiello, *The Dissipative Brain*, in Brain and Being - at the boundary between science, philosophy, language and arts, Globus G.G., Pribram, K.H. and Vitiello, G., (eds), John Benjamins Pub. Co. Amsterdam, 2004, 317-331).

The above explanations seem to furnish us with tangible information as regards what actually happens as regards memory and consciousness. Memory before is transits to another energy level in the quantum reality of the brain, occupies a vacuum state. When the information that comes from the environment is inserted in the brain, symmetry (SBS) is broken and there is a jump from the vacuum level to the first energy level and so on. The breaking of symmetry is to allow for order, because at this point in time, memory is activated, and the human person (brain), assumes one state from whence it operates. In the words of Vitiello, Symmetry "corresponds to indistinguishable points".[43] But the symmetry which gets broken in the creation of observable ordered patterns is the symmetry of the dynamical equations; symmetry is said to be spontaneously broken when the symmetry of the ground state is not the symmetry of the dynamical equations.[44] Once the external stimulus is intercepted by the body (brain), a necessary SBS occurs, a definite ground state, from the many infinite ground states or vacua is chosen. Massless NGB are observed, which carry long-range correlation waves observed in the brain dynamics. As a consequence of this, the time-reversal symmetry[45] is also broken, because the choice of a ground state, entails that the information has been recorded.[46] After information has been recorded, the brain state is fixated and the brain cannot be brought to the state configuration in which it was before the information printing occurred, for before the information recording process, the brain can in principle be in anyone of the infinitely many (unitarily inequivalent) vacua.[47] Time reversal symmetry means that the human person assumes a present state of affairs, one that can be distinguished from the past, and the future.

**Conclusion**

Quantum Interpretation of the Brain, with the research that is ongoing in the field of QBD, Quantum Information, Quantum Computing etc., is a process that is yet unfolding; it thus cannot be said that we have come to comprehensively decode what happens as regards consciousness, memory, information processing, learning and indeed the entire cognitive functions of the human brain. Yet what we are assured of is that with QBD, a quantum interpretation of the Brain has been born. This is further enhanced with the reality of Information processing in the brain, which is presently being explained fundamentally with the reality of QFT. What lies behind all these is that fact "quantum physics is a science of matter".[48] Before now any analysis of the brain, paid respect to Neurons, and today, even though brain cells are still very much needed to understanding brain tissues, yet we have attained to more depth in this understanding with the advent of QBD.

REFERENCES

---

[42] Ibid.
[43] G. Vitiello, *My Double Unveiled. The Dissipative Quantum Model of Brain*, John Benjamins, Amsterdam, 2001, 30
[44] Ibid, 30-31. According to Vitiello, «the word "spontaneous" means that the symmetry of the dynamics can be rearranged in any one of the possible ordering patterns observable at the physical level (in other words any of the physical phases can be dynamically realized).» (G. Vitiello, *My Double Unveiled*, 31)
[45] Time reversal symmetry means that the human person assumes a present state of affairs, one that can be distinguished from the past, and the future.
[46] G. Vitiello, *My Double Unveiled*, 107.
[47] Ibid.
[48] K.H. Pribram, *Brain and Mathematics* in Brain and Being - at the boundary between science, philosophy, language and arts, Globus G.G., Pribram, K.H. and Vitiello, G., (eds), John Benjamins Pub. Co. Amsterdam, 2004, 217-241.